\documentstyle[12pt]{article}
\begin{document}
\baselineskip=22pt
\begin{flushright}
OITS-546 \\
June 1994 (revised)
\end{flushright}
\vskip .5cm
\begin{center} EXCLUSIVE AND SEMI-INCLUSIVE B DECAYS \\
BASED ON \hspace{0.3cm} b $\rightarrow$ s \hspace{0.01cm}$\eta_c$
\hspace{0.3cm} TRANSITION

\vspace{0.5cm} N. G. DESHPANDE$^1$ and JOSIP TRAMPETIC$^2$

\vspace{0.5cm} {\it $^1$Institute of Theoretical Science, University of
Oregon, \\
 Eugene, OR 97403-5203, U.S.A.} \\ {\it $^2$Department of Theoretical
Physics, R. Bo\v{s}kovi\'{c}
 Institute, \\
 P.O.Box 1016, 41001 Zagreb, Croatia}
\end{center}

\vspace{1cm}

\begin{center}
ABSTRACT
\end{center}

\vspace{1cm} {\noindent We calculate the semi-inclusive processes B
$\rightarrow X_{s} \eta_c$ and relate it to B  $\rightarrow X_{s} \psi$ in
next to leading order QCD, and show how it can be used to determine an
accurate value of $f_{\eta_{c}}$. The exclusive modes B  $\rightarrow$
K$\eta_c$ and B  $\rightarrow$ K$\psi$ also yield similar results, whereas
B  $\rightarrow$ K$^*\eta_c$ and B  $\rightarrow$ K$^*\psi$ can provide
valuable information on hadronic form factors. We also estimate the
branching ratios and the ratios of exclusive to inclusive decay
modes dominated by $b \rightarrow s\eta_{c}$.}
\vspace{0.7cm}

\newpage

The b  $\rightarrow$ s$\eta_c$ transition offers a unique opportunity to
test our understanding of B meson decays. The related process  b
$\rightarrow$ s$\psi$ is known to give the ratio for semi-inclusive decays
of B, B  $\rightarrow$ $\psi +$ anything to exclusive decays B
$\rightarrow$ K$\psi$ and B $\rightarrow$ K$^*\psi$, in fair agreement
with data [1]. Here we show that  by taking the ratio of processes
involving $\eta_c$ to those involving $\psi$, one can remove the model
dependence to a large extent, and have an independent and powerful way of
determining $f_{\eta_{c}}$, the pseudoscalar decay constant of
$\eta_c$, the $S_0$ state of the charmonium.

The weak Hamiltonian, corrected to the next to leading order (NLO) in QCD,
relevant for us is given by [2,3]

\begin{equation}
 H_{\Delta B=1}^{\Delta S=-1} =
\frac{G_{F}}{\sqrt{2}}\left[\sum_{q=u,c}V_{qb}V_{qs}^{*}
\left(c_{1}O_{1}^{q}+c_{2}O_{2}^{q}\right)  - V_{tb}V_{ts}^{*}
\sum_{i=3}^{6}c_iO_i\right] + h.c.,
\end{equation} where $ c_{i}$ are the Wilson
coefficients and $O_{i}$ are the operators:

\begin{equation}
 O_{1}^{q} = \overline{s_{i}}\gamma^{\mu}\left(1-\gamma_{5}\right)b_{j}
 \overline{q_{j}}\gamma^{\mu}\left(1-\gamma_{5}\right)q_{i},
\hspace*{.5cm}
 O_{2}^{q} = \overline{s_{i}}\gamma^{\mu}\left(1-\gamma_{5}\right)b_{i}
 \overline{q_{j}}\gamma^{\mu}\left(1-\gamma_{5}\right)q_{j},
\end{equation}
\begin{equation}
 O_{3} = \overline{s_{i}}\gamma^{\mu}\left(1-\gamma_{5}\right)b_{i}
\sum_{q'}\overline{q'_{j}}\gamma_{\mu}\left(1-\gamma_{5}\right)q'_{j},
\hspace*{.5cm}
 O_{4} =\overline{s_{i}}\gamma^{\mu}\left(1-\gamma_{5}\right)b_{j}
 \sum_{q}\overline{q'_{j}}\gamma_{\mu}\left(1-\gamma_{5}\right)q'_{i},
\end{equation}
\begin{equation}
 O_{5} = \overline{s_{i}}\gamma^{\mu}\left(1-\gamma_{5}\right)b_{i}
\sum_{q'}\overline{q'_{j}}\gamma_{\mu}\left(1+\gamma_{5}\right)q'_{j},
\hspace*{.5cm}
 O_{6} = \overline{s_{i}}\gamma^{\mu}\left(1-\gamma_{5}\right)b_{j}
\sum_{q'}\overline{q'_{j}}\gamma_{\mu}\left(1+\gamma_{5}\right)q'_{i}.
\end{equation}

The values of the Wilson coefficients at the scale $\mu \cong m_{b}$, for
\begin{equation}
 m_{b} = 4.8 GeV,\hspace*{1cm} \Lambda_{\overline{M}S} = 250 MeV,
\hspace*{1cm} m_{top} = 150GeV,
\end{equation}
are [2]:
\newpage
\begin{eqnarray} c_{1} = 1.133,\hspace*{.1cm} c_{2} =-0.291,\hspace*{.1cm}
c_{3} = 0.015,
\nonumber \\ c_{4} =-0.034,\hspace*{.1cm} c_{5} =-0.010,\hspace*{.1cm}
c_{6} =-0.042.
\end{eqnarray}

We now define the matrix elements
\begin{equation}
 <0|\overline{c}\gamma_\mu c|\psi(q)>=i\varepsilon_\mu(q)g_\psi,
 \hspace*{1cm}
 <0|\overline{c}\gamma_\mu\gamma_5 c|\eta_c(q)>=iq_\mu f_{\eta_{c}},
\end {equation}
where $g^2_\psi=(1.414\pm0.083)$ GeV$^4$ from $\psi
\rightarrow e^+e^-$ [4]. The pseudoscalar decay constant $f_{\eta_{c}}$ is
expected to be $\approx$ 350 MeV, based on potential models, but the
exact value is not known.  The effective Hamiltonians in momentum space
for the two decays are
\begin{equation}
 H_{eff}^{b\rightarrow s\psi}=\frac{G_F}{\sqrt{2}}|V^*_{cs}V_{cb}|
|C_{\psi}|g_\psi\varepsilon_\psi^\mu(q)\overline{s_i}(k)
\gamma_\mu(1-\gamma_5)b_i(p),
\end{equation}
\begin{equation} H_{eff}^{b \rightarrow
s\eta_c}=\frac{G_F}{\sqrt{2}}|V^*_{cs}V_{cb}||C_{\eta_c}|f_{\eta_{c}}
q^\mu \overline{s_i}(k)
\gamma_\mu(1-\gamma_5)b_i(p),
\end{equation}
 where
\begin{equation} C_{\psi}=c_{2}+c_{3}+c_{5}+ \frac{1}{N_{c}}
\left(c_{1}+c_{4}+c_{6}\right),
\end{equation}
\begin{equation} C_{\eta_{c}}=c_{2}+c_{3}-c_{5}+ \frac{1}{N_{c}}
\left(c_{1}+c_{4}-c_{6}\right).
\end{equation}

We shall treat $C_{\psi}$ and $C_{\eta_{c}}$ as a phenomenological
parameters, thus absorbing in theirs definition any higher-order
correction or deviation from factorization that may arise [1].
The value of $|C_{\psi}|$ determined from
$BR$(B $\rightarrow X_{s} \psi$) =  $(1 \pm 0.25)\times10^{-2}$ [5], where
we have corrected for the fact that not all $\psi$ 's are produced directly
but arise from cascades of $\psi^\prime$ 's, etc., is

\begin{equation} |C_{\psi}|=0.220\pm0.026.
\end{equation}

\newpage

This value of $|C_{\psi}|$ favors $N_{c}\cong$2. We note that the ratio
$|C_{\eta_{c}}/C_{\psi}|$ is one in leading order.  In the NLO, with the
QCD coefficients from Ref's [2] and/or [3], for the
$2 \leq N_{c} \leq 3$ and m$_{t}$ = 174 $\pm$ 25 GeV
the ratio is stable within the following values:

\begin{equation}
|C_{\eta_c} / C_{\psi}| = 0.89 \pm 0.02.
\end{equation}

The ratio of semi-inclusive $\psi$ production to $\eta_c$
production can now be easily calculated. We find that
\begin{eqnarray}
\frac{\Gamma(B\rightarrow X_{s}\eta_c)}{\Gamma(B\rightarrow X_{s}\psi)} &
\equiv &  \frac{\Gamma(b\rightarrow s\eta_c)}{\Gamma(b\rightarrow s\psi)}
 = \left|f_{\eta_{c}}\frac{C_{\eta_{c}}}{C_{\psi}}\right|^{2}
\left(\frac{m_\psi}{g_\psi}\right)^2
    \left(\frac{\lambda^{b}_{s\eta_c}}
       {\lambda^{b}_{s\psi}}\right)^{1/2}  \nonumber \\
\vspace{1.0cm}    & \times &  \frac{[(m^2_b-m^2_s)^2 -m^2_{\eta_{c}}
      (m^2_b+m^2_s)]}{[m^2_b(m^2_b+m^2_\psi)-m^2_s(2m^2_b-m^2_\psi)
      +m^4_s-2m^4_\psi]} \nonumber \\
\vspace{1.0cm}    & \cong & 4.0({\rm GeV}^{-2})
\left|f_{\eta_{c}}\frac{C_{\eta_{c}}}{C_{\psi}}\right|^{2}
\end{eqnarray} where
\begin{eqnarray}
\lambda^{a}_{bc}=\left(1-\frac{m^2_{b}}{m^2_a}-
\frac{m^2_c}{m^2_a}\right)^2-4\frac{m^2_{b}m^2_c}{m^4_a}. \nonumber
\end{eqnarray} \\
Measurements of this hadron-model independent ratio offers a very
accurate determination of $f_{\eta_{c}}$. In ratio (14) , again only  the
direct $\eta_c$ production should be included. Since the cascade of
charmonia into $\eta_c$ is extremely small, no correction is necessary.

Next we consider the B $\rightarrow K\psi $ and B $\rightarrow K\eta_c $ modes.
Using the general Lorentz decomposition of the matrix element
\begin{equation} <K(k)|\overline{s}\gamma_\mu b|B(p)>=(p+k)_\mu
f^{(+)}_{KB}(q^2)+
  q_\mu f^{(-)}_{KB}(q^2),
\end{equation}
\newpage
we have
\begin{eqnarray}
\frac{\Gamma(B\rightarrow K\eta_c)}{\Gamma(B\rightarrow K\psi)} & = &
\left|f_{\eta_{c}}\frac{C_{\eta_{c}}}{C_{\psi}}\right|^{2}
\left(\frac{m_\psi}{g_\psi}\right)^2
  \frac{(\lambda^{B}_{K\eta_{c}})^{1/2}}{(\lambda^{B}_{K\psi})^{3/2}}
  \left|f^{(+)}_{KB}(m^2_{\psi})\right|^{-2} \nonumber \\
 & \times & \left|\left(
1-\frac{m^2_K}{m^2_B}\right)f^{(+)}_{KB}(m^2_{\eta_{c}}) +
 \frac{m^2_{\eta_{c}}}{m^2_B}f^{(-)}_{KB}(m^2_{\eta_{c}})\right|^2.
\end{eqnarray}
Since $m_\psi \cong m_{\eta_{c}}$, we have set
$ f^{(+)}_{KB}(m^2_{\eta_{c}})/f^{(+)}_{KB}(m^2_{\psi}) \cong 1 $. The
second term in Eq.16 is
$\left[m^{2}_{\eta_{c}}/m^{2}_{B}\right]
\left[ f^{(-)}_{KB}(m^2_{\eta_{c}}) /f^{(+)}_{KB}(m^2_{\eta_{c}})\right]
\cong -0.06 $, and it is negligible compared with the first term in (16),
in accord with the pole approximation as discussed in [6]. Because of that,
essentially hadron-model independent, ratio (16) is:
\begin{equation}
\Gamma(B\rightarrow K\eta_c)/\Gamma(B\rightarrow K\psi) \cong
14.2({\rm GeV}^{-2})\left|f_{\eta_{c}}\frac{C_{\eta_{c}}}
{C_{\psi}}\right|^{2}.
\end{equation}

Finaly, we calculate the exclusive rates for B $\rightarrow K^{*}\psi$ and
B  $\rightarrow K^{*}\eta_c$.
The general Lorentz decomposition of the (B$\rightarrow$ K$^{*}$) matrix
element:
\begin{eqnarray}
 & <K^*(k)|\overline{s}\gamma_\mu(1-\gamma_5)b|B(p)>  =
    -i\varepsilon_{\mu\nu\lambda\sigma}(p+k)^\nu(p-k)^\lambda
   \varepsilon^\sigma(k)V      \nonumber \\
 & +\varepsilon(k)(m^2_B-m^2_{K^*})A_1-(q \cdot \varepsilon (k))
      (p+k)_\mu A_2  \\
 & +(q \cdot \varepsilon (k))(m_{B}+m_{K^{*}})
    (q_\mu/q^2)\left[2m_{K^{*}}A_0-(m_B-m_{K^{*}})(A_1-A_2)\right], \nonumber
\end{eqnarray}
and the corresponding definitions of relevant form-factors

\begin{equation}
V  =  \frac{V\left(q^{2}\right)}{(m_{B}+m_{K^{*}})},
\hspace*{.5cm}
V\left(q^{2}\right)=\frac{V(0)}{\left(1-q^{2}/m_{1^{-}}^{2}\right)}
\end{equation}

\begin{equation}
A_{0}  =  \frac{A_{0}\left(q^{2}\right)}{(m_{B}+m_{K^{*}})},
\hspace*{.5cm}
A_{0}\left(q^{2}\right)=\frac{A_{0}(0)}{\left(1-q^{2}/m_{0^{-}}^{2}\right)}
\end{equation}

\begin{equation}
A_{1}  =  \frac{A_{1}\left(q^{2}\right)}{(m_{B}-m_{K^{*}})},
\hspace*{.5cm}
A_{1}\left(q^{2}\right)=\frac{A_{1}(0)}{\left(1-q^{2}/m_{1^{+}}^{2}\right)}
\end{equation}

\begin{equation}
A_{2}  =  \frac{A_{2}\left(q^{2}\right)}{(m_{B}+m_{K^{*}})},
 \hspace*{.5cm}
A_{2}\left(q^{2}\right)=\frac{A_{2}(0)}{\left(1-q^{2}/m_{1^{+}}^{2}\right)}
\end{equation}

\begin{equation}
A_{0}(0) = \frac{m_{B}+m_{K^{*}}}{2m_{K^{*}}}A_{1}(0)
- \frac{m_{B}-m_{K^{*}}}{2m_{K^{*}}}A_{2}(0),
\end{equation}

gives the hadron-model dependent ratio
\begin{eqnarray}
\frac{\Gamma(B \rightarrow K^*\eta_c)}{\Gamma(B \rightarrow K^*\psi)}
&  = & \left|f_{\eta_{c}}\frac{C_{\eta_{c}}}{C_{\psi}}\right|^{2}
\left(\frac{m_B+m_{K^*}}{g_\psi}\right)^2
     \left(\frac{\lambda^{B}_{K^{*}\eta_{c}}}
  {\lambda^{B}_{K^{*}\psi}}\right)^{3/2}
   \left|A_0\right|^2 \\
& \times & \left[2\left|V\right|^{2}
       +\left(\frac{3}{\lambda^{B}_{K^{*}\psi}}
       +\frac{m^4_B}{4m^2_{K^{*}}m^2_\psi}\right)\left(1-\frac{m^2_{K^{*}}}
       {m^2_B}\right)^2\left|A_1\right|^2+\right. \nonumber \\
 & + &
\left(\frac{m^4_B}{4m^2_{K^{*}}m^2_\psi}\right)\lambda^{B}_{K^{*}\psi}
         \left|A_2\right|^2 -   \nonumber \\
& - & \left. \left(\frac{m^4_B}{2m^2_{K^{*}}m^2_\psi}\right)\left(1-\frac
       {m^2_\psi}{m^2_B}-\frac{m^2_{K^{*}}}{m^2_B}\right)\left(
        1-\frac{m^2_{K^{*}}}{m^2_B}\right)A_1A_2\right]^{-1}. \nonumber
\end{eqnarray}

The ratio (24) can be represented as

\begin{equation}
\Gamma(B\rightarrow K^{*}\eta_c)/\Gamma(B\rightarrow K^{*}\psi) =
  R ({\rm GeV}^{-2})\left|f_{\eta_{c}}\frac{C_{\eta_{c}}}
{C_{\psi}}\right|^{2}.
\end{equation}
The factor R depends critically on the form factors. We evaluate it in three
models for illustration. The first is the relativistic quark model, Model
1, [7].
For Model 2, we use a parametrization based on effective chiral theory for
mesons with flavor and spin symmetries of Heavy Quark Effective Theory
[HQET] [8,9].
This method contains errors that follow from semileptonic D-meson decays
used as
inputs [10]. The Model 3 is a variation [11], where $A_{2}(0)$ is set equal
to zero
and V(0) is renormalized from the value in Ref.[7]. Last model gives a much
better
fit to experimental measurements of $\Gamma_{L}/\Gamma_{T+L}$ ratio,
where $\Gamma_{L(T)}$ refers to the
longitudinal(transvers) polarization of $\psi$ mesons in $B \rightarrow
K^*\psi$
decay. This ratio is also very sensitive to the form factors. All models
use the single pole fits for V and $A_{i}$, (i=0,1,2) form-factors.
The masses of excited meson states, corresponding to the
$\overline{s}b$ current, are:
$m_{0^{-}}$ = 5.38, $m_{1^{-}}$ = 5.43, $m_{0^{+}}$ = 5.89 and
$m_{1^{+}}$ = 5.82, GeV's.  In Table 1 we summarise the parameters (V,A$_{i}$)
and the factor R for $B \rightarrow K^*(\psi,\eta_{c})$ decays, while
the ratios $\Gamma_{L}/\Gamma_{T+L}$, $\Gamma_{T}/\Gamma_{T+L}$
are given for $B \rightarrow K^*\psi$ decay.\\

\centerline{\it Table 1 }
\vspace{.01cm}
\noindent
The (V,A$_{i}$), R,  $\Gamma_{L}/\Gamma_{T+L}$ and $\Gamma_{T}/\Gamma_{T+L}$
in Models 1, 2, 3 and 1$^\prime$, 2$^\prime$.
\vspace{.5cm}\\
\begin{tabular}{|c|c|c|c|c|c|c|c|} \hline \hline
 & & & & & & & \\
Model&V(0)&$A_{0}(0)$&$A_{1}(0)$&$A_{2}(0)$& R
& $\Gamma_{L}/\Gamma_{T+L}$ & $\Gamma_{T}/\Gamma_{T+L}$ \\
$[Ref.]$ & & & & & ($GeV^{-2}$) & & \\
\hline \hline
 & & & & & & & \\
1 [7] &0.369 &0.321 &0.328 &0.331 &1.01 &0.573 &0.427 \\
 & & & & & & & \\

2 [8,9] &0.61 &0.20 &0.20 &0.20 &0.67 &0.366 &0.634 \\
 & & & & & & & \\

3 [11] &0.656 &1.133 &0.328 &0.0 &5.49 &0.728 &0.272 \\
 & & & & & & & \\
\hline
 & & & & & & & \\
1$^\prime$ &0.37 &1.13 &0.33 &0.0 &6.0 &0.80 &0.20 \\
 & & & & & & & \\

2$^\prime$ &0.61 &0.69 &0.20 &0.0 &4.7 &0.63 &0.37 \\
 & & & & & & & \\
\hline \hline
\end{tabular} \\

\vspace{0.7cm}
The experimental value of  $\Gamma_{L}/\Gamma_{T+L}$ is:
\begin{equation}
\frac{\Gamma_{L}(B \rightarrow K^*\psi)}{\Gamma_{T+L}(B \rightarrow K^*\psi)}=
\left \{ \begin{array}{ll}
0.80 \pm 0.05^{+0.08}_{-0.08} &, [12]\\
\\
0.66 \pm 0.10^{+0.08}_{-0.10} &. [13]\\
\end{array}
\right.
\end{equation}
Note that if the condition $A_{2}(0)=0$ is put by hand in Models 1 and 2,
we find
$\Gamma_{L}/\Gamma_{T+L} \cong $ 0.80 and 0.63, respectively. In Table 1 we
call
these models 1$^\prime$ and 2$^\prime$.
Measurement of $D \rightarrow K^* l \nu$ decay [10]
favors $A_{2}(0)\cong 0$, and HQET relates ($B \rightarrow K^*$) to
($D \rightarrow K^*$) transition  [11].

As a by-product of the above analysis and using recent CLEO II experimental
results on the $ B \rightarrow K^{*}\psi $ and  $ B \rightarrow K\psi $
decays [12]:

\begin{equation} BR(B^{-} \rightarrow K^{*-} \psi) = (0.178 \pm 0.051 \pm
0.023) \times 10^{-2} ,
\end{equation}
\begin{equation} BR(B^{0} \rightarrow K^{*0} \psi) = (0.169 \pm 0.031 \pm
0.018) \times 10^{-2} ,
\end{equation}
\begin{equation} BR(B^{-} \rightarrow K^{-} \psi) = (0.110 \pm 0.015 \pm
0.009) \times 10^{-2} ,
\end{equation}
\begin{equation} BR(B^{0} \rightarrow K^{0} \psi) = (0.075 \pm 0.024 \pm
0.008) \times 10^{-2} ,
\end{equation}
from the expressions,

\begin{equation}
\frac{\Gamma(B \rightarrow X_{s} \eta_{c})}{\Gamma(B \rightarrow X_{s}
\psi)} = \frac{4.0}{R}
\frac{\Gamma(B \rightarrow K^{*} \eta_{c})}{\Gamma(B \rightarrow K^{*}
\psi)},
\end{equation}
\begin{equation}
\frac{\Gamma(B \rightarrow X_{s} \eta_{c})}{\Gamma(B \rightarrow X_{s}
\psi)} = \frac{4.0}{14.2}
\frac{\Gamma(B \rightarrow K \eta_{c})}{\Gamma(B \rightarrow K \psi)},
\end{equation}
we estimate the ratios of exclusive to inclusive decay
modes:
\begin{equation}
\Gamma(B^{-} \rightarrow K^{*-} \eta_{c}) /
\Gamma(B \rightarrow X_{s} \eta_{c}) = R (0.045 \pm 0.018) ,
\end{equation}
\begin{equation}
\Gamma(B^{0} \rightarrow K^{*0} \eta_{c}) /
\Gamma(B \rightarrow X_{s} \eta_{c}) = R (0.042 \pm 0.014),
\end{equation}
\begin{equation}
\Gamma(B^{-} \rightarrow K^{-} \eta_{c}) /
\Gamma(B \rightarrow X_{s} \eta_{c}) = 0.391 \pm 0.116,
\end{equation}
\begin{equation}
\Gamma(B^{0} \rightarrow K^{0} \eta_{c}) /
\Gamma(B \rightarrow X_{s} \eta_{c}) = 0.266 \pm 0.112.
\end{equation}

The above estimates reflect the experimental errors from (27) to (30).
Theoretically, the ratios in (33) and (34) have to be equal,
as also (35) and (36).
If one uses the values of R from models based on A$_2$(0)=0, then
$\Gamma(B \rightarrow K^{*} \eta_{c})$ is $\cong$ 25\%
of the inclusive rate. This seems to be surprisingly large because it
would imply that $B\rightarrow K\eta_{c}$ and $B\rightarrow K^{*}\eta_{c}$
predominantely saturate transition $B \rightarrow X_{s} \eta_{c}$ !

To estimate branching ratios for  B $\rightarrow$ K$^*\eta_{c}$,
B $\rightarrow$ K $\eta_{c}$ and  B $\rightarrow X_{s} \eta_c$ decays one
has to know the pseudoscalar decay constant $f_{\eta_{c}}$.
Theoretically, like the value of $g_\psi$, the value of $f_{\eta_{c}}$
can be related to the wave function of the $S$-state of the charmonium at
the origin:
\begin{equation}
 g_{\psi}^{2} = 12m_{\psi}|\psi(0)|^{2},
\end{equation}
\begin{equation} f_{\eta_{c}}^{2} =
48\frac{m_c^2}{m_{\eta_{c}}^3}|\psi(0)|^{2}.
\end{equation}
Without QCD corrections the expressions (37,38) gives $f_{\eta_{c}} \cong $ 350
MeV. The QCD corrections are significant but approximately cancel in
the ratio [14], and one expects $f_{\eta_{c}}$ to be close to the above value.

A non-perturbative estimate of $f_{\eta_{c}}$ based on QCD sum
rules [15] could be more reliable. Saturating vacuum polarization
of pseudoscalar density evaluated to $O(\alpha_{s})$ with $m_{\eta_{c}}$,
$m_{\eta^{'}_{c}}$ and using n=4,5 moments, which minimize the
continuum contributions, we find
\begin{equation}
f_{\eta_{c}} = (300 \pm 50) MeV.
\end{equation}
The major source of error above is not
$\alpha_{s}(m_{\eta_{c}}) = 0.212 \pm 0.021$, but the current quark mass of
charm which we take as $m_{c} = (1.25 \pm 0.03)GeV$.

Using the central value of $f_{\eta_{c}}$ = 300 MeV,
and taking the ratio $|C_{\eta_c}/C_{\psi}|$ = 0.89, we estimate
branching ratios for B $\rightarrow$ K$^*\eta_{c}$, K $\eta_{c}$
and $X_{s}\eta_c$ decays:
\begin{equation} BR(B^{-} \rightarrow K^{*-}\eta_{c}) = R (0.127 \pm 0.040)
\times 10^{-3} ,
\end{equation}
\begin{equation} BR(B^{0} \rightarrow K^{*0}\eta_{c}) = R (0.121 \pm 0.026)
\times 10^{-3} ,
\end{equation}
\begin{equation} BR(B^{-} \rightarrow K^{-}\eta_{c}) = (1.113 \pm 0.177)
\times 10^{-3} ,
\end{equation}
\begin{equation} BR(B^{0} \rightarrow K^{0}\eta_{c}) = (0.759 \pm 0.256)
\times 10^{-3} ,
\end{equation}
\begin{equation} BR(B\rightarrow X_{s} \eta_c) = (2.851 \pm 0.713)
\times 10^{-3} .
\end{equation}

To conclude, we have shown a very accurate technique of measuring
 $f_{\eta_{c}}$. The consistency of this value obtained from
semi-inclusive decays and exclusive decays will check factorization.
Once $f_{\eta_{c}}$ is known from the measurements of B $\rightarrow
X_{s} \eta_c$,  this can be used to see if the form-factor ratio
(24) is consistent with models of those decays.
The measurement of B $\rightarrow$ K$^*\eta_c$ probes the spin-0 part of the
axial form factor, and provides a useful check on the model building.
Finaly, we estimated branching ratios and the ratios of exclusive to
inclusive decays for the most interesting modes.

\vspace{1cm} This work was supported in part by the U.S. Department of
Energy under Contract DE-FG06-85ER40224, and by the Croatian Ministry of
Research under Contract 1-03-199.


\newpage

\begin{thebibliography}{99}

\bibitem{}  K. Berkelman, "Hadronic Decays - Experiment",\\
            p.136 in B-Decays, ed. by: S. Stone, World Scientific, 1992;

            N.G. Deshpande, J. Trampeti\'c, and K. Panose,\\
            Phys. Lett. B214, 467 (1988).

\bibitem{}  A.J. Buras, M. Jamin, M.E. Lautenbacher and P.H. Weisz,\\
            Nucl. Phys. B370, 69 (1992).

\bibitem{}  N.G. Deshpande, Xiao-Gang He, OITS-538 preprint (1994)

\bibitem{}  Review of Particle Properties, Phys. Rev. D45, II.18 (1992).

\bibitem{}  N.G. Deshpande, J. Trampetic and K. Panose,

            Phys. Lett. B214, 467 (1988).

\bibitem{}  J.G. K\"orner, K. Schilcher, M. Wilber and Y.L. Wu,

            Z.Phys. C48, 663 (1990).

\bibitem{}  M. Wirbel, B. Stech and M. Bauer, Z.Phys. C29, 637 (1985),

            M. Bauer, B. Stech and M. Wirbel, Z.Phys. C34, 103 (1987).

\bibitem{}  R. Casalbuoni, A. Deandrea,  N. Di Bartolomeo, F. Feruglio,

             R. Gatto and G. Narduli, Phys. Lett. B299, 139 (1993) .

\bibitem{}  A. Deandrea, N. Di Bartolomeo, R. Gatto and  G. Narduli,\\
            Phys. Lett. B318, 549 (1993).

\bibitem{}  J.C. Anjos et al., Phys. Rev. Lett. 65, 2630 (1990).

\bibitem{}  G. Kramer, T. Mannel and G.A. Schuler,  Z.Phys. C51, 649 (1991)

\bibitem{}  T.E. Browder, K. Honscheid and S. Playfer, \\
            "A Review of Hadronic and Rare B Decays",\\
            (To appear in B-Decays, 2nd edition, Ed. by S. Stone,
            World Scientific);\\
            CLEO Collaboration, M.S. Alam et al., CLNS 94-1270, CLEO 94-5.

\bibitem{}  CDF Collaboration, FERMILAB-CONF-94/127-E

\bibitem{}  W. Kwong, J.L. Rosner and C. Quigg,

            Ann. Rev. Nucl. Part. Sci. 37, 325-382 (1987).

\bibitem{}  L.J. Reinders, H.R. Rubinstein and S. Yazaki,\\
            Phys. Lett. B94, 203 (1980);\\
            L.J. Reinders, H.R. Rubinstein and S. Yazaki,\\
            Phys. Reports 127, 1-97 (1985).

\end{thebibliography}
\end{document}